\documentclass[aps,prl,twocolumn,preprintnumbers,floatfix,reprint]{revtex4-1}
\usepackage{mathrsfs,natbib}
\usepackage{amsmath,amssymb, amsfonts, bbm}
\usepackage{graphics,epsfig,color,subfigure}

\newcommand{\be}{\begin{equation}}
\newcommand{\ee}{\end{equation}}

\def\bea{\begin{align}}
\def\ena{\end{align}}

\def\beqa{\begin{eqnarray}}
\def\enqa{\end{eqnarray}}

\newcommand{\bx}{\mbox{\boldmath $x$}}
\newcommand{\bfe}{\mbox{\boldmath $e$}}

\begin{document}

\title{Calculation of the One $W$ Loop $H\to \gamma \gamma$ Decay Amplitude with a Lattice Regulator}

\author{Francis Bursa$^{1}$}
\email{f.bursa@swansea.ac.uk}
\author{Aleksey Cherman$^{2}$}
\email{a.cherman@damtp.cam.ac.uk}
\author{Thomas C. Hammant$^{2}$}
\email{tch36@damtp.cam.ac.uk}
\author{Ron R. Horgan$^{2}$}
\email{r.r.horgan@damtp.cam.ac.uk}
\author{Matthew Wingate$^{2}$}
\email{m.wingate@damtp.cam.ac.uk}
\affiliation{
$^1$ College of Science, Swansea University, Singleton Park, Swansea SA2 8PP, United Kingdom \\
$^2$ Department of Applied Mathematics and Theoretical Physics, University of Cambridge, Centre for Mathematical Sciences, Cambridge CB3 0WA, United Kingdom
}

\preprint{DAMTP-2011-107}

\begin{abstract}
  There has been a controversial recent claim that the standard result on the
  Higgs to two photon decay rate is incorrect, with the use of dimensional
  regularization fingered as the alleged culprit.  Given the great importance
  of the $H\to \gamma \gamma$ process as a possible Standard Model Higgs
  discovery channel at the LHC if the Higgs mass is light, it is critical to
  find a way to check the correctness of the results of dimensional
  regularization for this process.  Here we report the results of a
  perturbative calculation of the $H\to \gamma \gamma$ decay amplitude using a
  spacetime lattice as a UV regulator, which is the only known gauge-invariant
  regulator for non-Abelian gauge theories other than dimensional
  regularization.  We find that the decay amplitude calculated using
  lattice-regularized perturbation theory is consistent to very high
  statistical accuracy with the decay amplitude obtained using dimensional
  regularization.
\end{abstract}

\maketitle

The decay of the Higgs boson into two photons, $H\to \gamma \gamma$, is a
crucial channel in the search for light Higgs bosons at the Large Hadron
Collider (LHC) (see e.g.~\cite{Djouadi:2005gi}) especially if the Higgs mass
is close to the LEP bound of $M_{H} \gtrsim
114.4~\mathrm{GeV}/c^2$\cite{Barate:2003sz}. In the Standard Model (SM) there
is no direct vertex for this decay, and the process proceeds through loops of
intermediate gauge bosons and quarks.  At one-loop level, the two most
important contributions to the decay amplitude are from the top quark loop and
from $W^{\pm}$ boson loops. The evaluation of the latter contribution has
recently been questioned.  The $W^{\pm}$-loop contribution to the $H\to \gamma
\gamma$ decay amplitude takes the form
\begin{equation}
\mathcal{M_{\mu \nu}} = -\frac{e^{2}g}{8\pi^{2}M_{W}} \left[k^{1}_{\mu}
  k^{2}_{\nu} - g_{\mu \nu} (k^{1} \cdot k^{2}) \right] \frac{F(\tau)}{\tau} \,,
\end{equation}
where $e$ and $g$ are the $U(1)_{EM}$ and $SU(2)_{L}$ couplings respectively,
$M_{W}$ is the $W$-boson mass, $k^{1,2}_{\mu}$ are the four-momenta of the
emitted photons, $\tau = M_{H}^{2}/4M_{W}^{2}$, and $M_{H}$ is the mass of the
Higgs boson.  The function $F(\tau)$ was first calculated in
\cite{Ellis:1975ap,Shifman:1979eb} many years ago, with the result
\begin{align}
\label{dimregamp}
F(\tau)  = \frac{3}{2} + \tau + \frac{3}{2}(2 - \tau^{-1}) 
[\arcsin \sqrt{\tau}]^{2} \,,
\end{align}
for $\tau \le 1$. These calculations used dimensional regularization, although
\cite{Shifman:1979eb} also derived some beautiful low-energy theorems on the
properties of the amplitude which did not explicitly call on any particular
regularization procedure.

Recently this amplitude has been examined again by Gastmans, Wu, and
Wu~\cite{Gastmans:2011ks,*Gastmans:2011wh}.
Ref.~\cite{Gastmans:2011ks,*Gastmans:2011wh} carried out the calculation in
unitary gauge directly in four dimensions without any use of dimensional
regularization; they used a particular momentum-routing prescription to handle
the divergences arising at intermediate stages in the calculation of the
amplitude.  The result they obtained is
\begin{equation}
\label{Gastmansamp}
F(\tau)  = \frac{3}{2} + \frac{3}{2}(2 - \tau^{-1}) [\arcsin \sqrt{\tau}]^{2} ,
\end{equation}
which differs from the classic result by the absence of a term linear in
$\tau$. This is a highly surprising result, since one expects that the
calculation of a physical amplitude in quantum field theory must not depend on
the regularization prescription.  This difference was blamed on a problem with
dimensional regularization.  The claim was that dimensional regularization
assigns a finite value to a contribution to the amplitude which is zero in
$d=4$ when evaluated using the prescriptions of
\cite{Gastmans:2011ks,*Gastmans:2011wh}.  If the striking result of
\cite{Gastmans:2011ks,*Gastmans:2011wh} were the correct form of the decay
amplitude, it would throw into question a vast number of calculations carried
out with dimensional regularization.  Furthermore, Eq.~(\ref{Gastmansamp})
gives a significantly smaller decay rate than Eq.~(\ref{dimregamp}) for Higgs
masses where the diphoton decay is an important discovery channel.  For
example, for $M_H=115 \mathrm{GeV}/c^2$ the decay width would be reduced by
$\approx50$\% \cite{Gastmans:2011ks,*Gastmans:2011wh}. If correct, this would
have significant implications for Higgs searches at the LHC.

Given the long experience with the reliability of dimensional regularization,
the claim of \cite{Gastmans:2011ks,Gastmans:2011wh} must naturally be viewed
with skepticism, and a number of papers\cite{Shifman:2011ri, *Huang:2011yf,
  *Marciano:2011gm, *Jegerlehner:2011jm,*Shao:2011wx,*Liang:2011sj} have
appeared criticizing the results of Gastmans \textit{et al.}\ from a variety
of perspectives.  However, given the extraordinary phenomenological importance
of correctly determining this amplitude in the Standard Model, it is highly
desirable to add an extra check by explicitly computing $M_{\mu \nu}$ using a reliable gauge-invariant
regulator other than dimensional regularization.  

{\it Lattice as a regulator} --- The list of reliable gauge-invariant
regulators for non-Abelian gauge theories is very short.  Naive momentum
cutoffs break gauge invariance, while Pauli-Villars-like regulators have
serious problems in non-Abelian theories (see e.g.\ \cite{Leon:1995nm}).
In perturbative gauge theory calculations dimensional
regularization\cite{`tHooft:1972fi} is omnipresent since it manifestly
preserves gauge invariance and is easy to use. The only other known
gauge-invariant regulator is a spacetime lattice, which has the effect of
imposing a momentum cutoff $\sim 1/a$, where $a$ is the lattice spacing, but
does this in a far more subtle way than a naive momentum cutoff, with manifest
gauge invariance at any $a$. Lattice regularization has the great advantage
that it can be used beyond perturbation theory, providing a nonperturbative
definition of a theory.  
This has led to much progress in the nonperturbative understanding of
gauge theories, e.g.\ using Monte Carlo methods to compute
the full lattice-regularized Euclidean path integral of QCD.  A lattice
regulator can also be used in perturbative calculations, where it gives an
alternative route to obtaining results that otherwise could only be obtained
using dimensional regularization.

Lattice regulators have two chief disadvantages for perturbative calculations
compared to dimensional regularization.  The first disadvantage is that a
lattice regulator breaks rotational and translation invariance, keeping only
discrete subgroups of these symmetries.
Rotational symmetry is crucial for enabling analytic evaluation of loop
integrals in perturbative calculations; the loss of these symmetries on the
lattice at finite $a$ means that loop integrals must be evaluated
numerically. 
This is certainly a practical inconvenience, but it is not a problem of
principle.  The second disadvantage is of a deeper nature.  It is not known
how to implement lattice regulators for chiral gauge theories; for
some recent reviews see
\cite{Luscher:2000hn,*Golterman:2000hr,*Poppitz:2010at}.  Since the Standard
Model is a chiral gauge theory this precludes the use of lattice regulators
for computing most SM observables sensitive to the electroweak sector.

Fortunately the difficulties with chiral fermion couplings do not play a role
here.  This is because only the $W^{\pm}$-mediated contribution to the $H\to
\gamma \gamma$ decay amplitude has been questioned, not the top-quark-mediated
contribution.  Using the lattice regulator we can simply compute the $H \to
\gamma \gamma$ decay amplitude with all the fermion fields 
turned off.

{\it Setup} --- We will carry out our calculations below the threshold for the
$H \to W^+W^-$ decay ($\tau<1$) since the LHC has already excluded SM Higgs
boson masses $m_{H}\gtrsim 146$ GeV\cite{LP2011:CMS,*LP2011:ATLAS} at $> 95\%$
confidence.  This assumption makes the Wick rotation to Euclidean space
especially simple. (Above threshold, the calculation could also be done using
lattice perturbation theory \cite{Hart:2006ij}.)  To evaluate the decay
amplitude, we go to Euclidean space, choose $\mu=\nu=1$, and work in the Higgs
boson rest frame, so that $k_1=(iM_H/2,0,0,M_H/2)$, $k_2=(iM_H/2,0,0,-M_H/2)$
for on-shell photons. The amplitude then takes the form
\begin{equation}
{\cal M}_{11}=-\frac{e^2gM_W}{4\pi^2}F(\tau) \,.
\end{equation}

The lattice calculation proceeds by calculating $F(\tau)$ numerically at a
range of values of the lattice spacing $a$, taking the continuum limit
$M_{W}a, M_{H}a\to 0$ with $\tau$ held fixed.
With this end in view we calculate the
dimensionless function $F(\tau,aM_W)$, where the dependence on $aM_W$ is due
to lattice artifacts, and perform a two dimensional fit in $\tau$ and $aM_W$
to recover continuum limit, $aM_W \to 0$, giving $F(\tau) \equiv F(\tau,0)$.
For the fit function we take as our guide the functional form obtained from
current analytic calculations and find it sufficient to consider
\begin{equation}
F(\tau,aM_W) = c_{1}+c_{2}\tau+c_{3}\left(2-\frac{1}{\tau}\right)[\arcsin{\sqrt{\tau}}]^{2} \,, 
\end{equation}
where $c_{1},c_{2},c_{3}$ are functions of $a M_{W}$.  

For the gauge boson action that we use, a simple parametrization of the $aM_W$
dependence of $c_1(aM_W)$ is found to be sufficient and $c_2$ and $c_3$ appear
constant for the values of $aM_W$ that we choose.

{\it Lattice Feynman rules} --- As in any perturbative calculation of an
amplitude, we must fix a gauge to proceed, and we choose to use unitary gauge.
The lattice action for the gauge and Higgs sector of Standard Model in unitary gauge is described in
Ref.~\cite{Montvay:1994cy}. To carry out the calculation we use the
\textsc{HiPPy} and \textsc{HPsrc} packages developed for automated lattice
perturbation theory \cite{Hart:2004bd,Hart:2009nr}. We use the
Symanzik-improved lattice action \cite{Luscher:1985zq} for the vector-boson
vertices that arise from the pure $SU(2)$ gauge part of the action.
The improvement is important since this greatly reduces lattice artifact
contributions.
These vertices are available in the automated packages and
reduce to the continuum ones in the $a \to 0$ limit. For the $W$
boson propagator we add a mass term to the quadratic part of the gauge boson
action and invert numerically to give the lattice version of the Proca
propagator, as is appropriate to unitary gauge.

\begin{figure}
\includegraphics[width=0.35\linewidth]{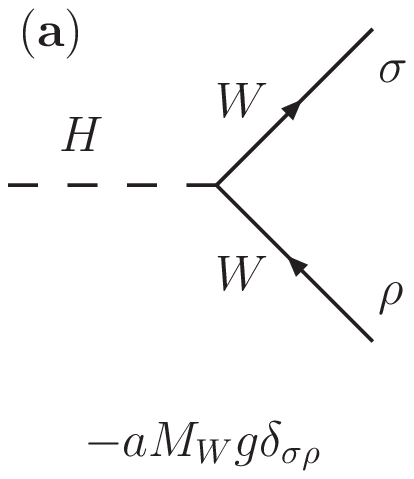}\;
\includegraphics[width=0.35\linewidth]{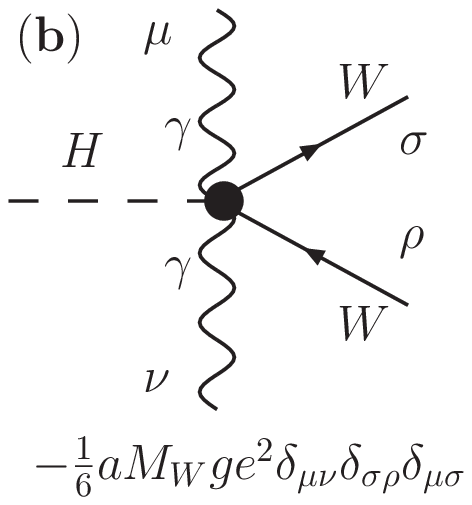}\\
\includegraphics[width=0.35\linewidth]{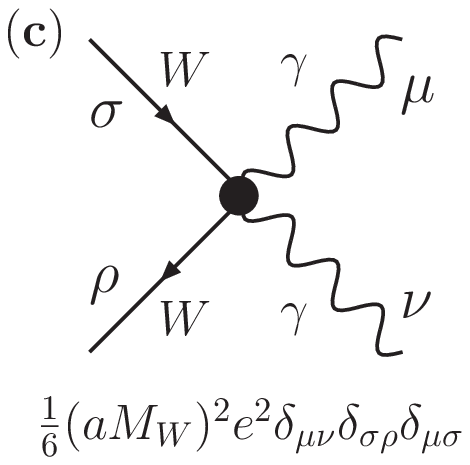}\;
\includegraphics[width=0.36\linewidth]{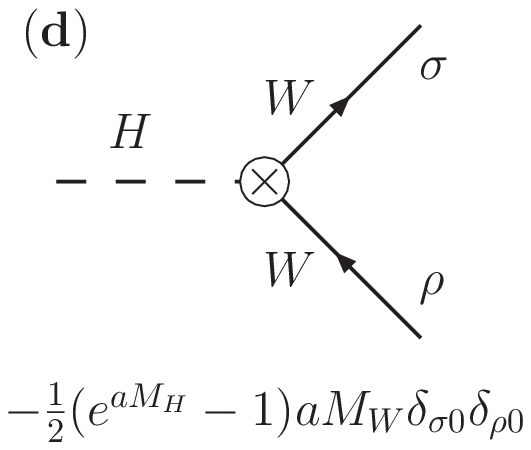}\;
\caption{\label{fig:Frules}Feynman rules derived from $L_H$ (\ref{eq:LH}).}
\end{figure}

The vector-boson mass term arises from the lattice Lagrangian
\cite{Montvay:1994cy}
\begin{align}
L_H = &-\frac{1}{2}\sum_{\mu=1}^{4}[av+aH(\bx+a\bfe_\mu)][av+aH(\bx)]\nonumber \\
&\times\mathrm{Tr}[e^{gaT_iW_{i\mu}(\bx)}e^{-g^{\prime}aT_3B_{\mu}(\bx)}]
\label{eq:LH}
\end{align}
where $v$ is the vacuum expectation value of the radial scalar field, $H(\bx)$
is the Higgs field, $W_i(\bx),~i=1,2,3$ are the $SU(2)_L$ gauge boson fields and
$B(\bx)$ is the $U(1)_Y$ gauge field. The $T_i$ are the anti-hermitian
generators of $SU(2)$ satisfying $[T_i,T_j] = -\epsilon_{ijk}T_k$. As is usual
for lattice actions $L_H$ is dimensionless. The term quadratic in the $W$ and
$B$ fields generates the $W^\pm$ and $Z$ mass terms with $M_W=M_Z = gv/2$ and
where
\begin{equation}
B_\mu = s_W Z_\mu + c_W A_\mu\;,~~~{W_3}_\mu = -c_W Z_\mu + s_W A_\mu\;. \label{BW3}
\end{equation} 
Here $s_W \equiv \sin\theta_W,~c_W \equiv \cos\theta_W$, where $\theta_W$ is
the Weinberg angle, $\tan\theta_W = g^\prime/g$. This term also contains the
usual $HW^+W^-$ vertex and, because the lattice spacing is non-zero, it
additionally gives rise to extra interaction terms on the lattice which have
no continuum counterpart. It should be emphasized that when inserted into a
Feynman diagram these extra terms give contributions which are non-zero in the
continuum limit; indeed, such contributions are a necessary consequence of
maintaining gauge invariance while regularizing the theory by restricting
momenta to lie in the lattice Brillouin zone. These extra Feynman rules are
derived by expanding the trace term in $L_H$, using the
Baker-Campbell-Hausdorff formula to the appropriate order and using
(\ref{BW3}) to identify the photon field $A_\mu$. The $HW^+W^-$ and additional
Feynman rules are displayed graphically in Fig.~\ref{fig:Frules}. The rule in
(d) arises because of the presence of $H(\bx+a\bfe_\mu)$ in $L_H$, and,
because the Higgs boson is at rest, the Lorentz index on the $W$-bosons
attached to this vertex is restricted to $\mu=0$ as shown in the figure.

\begin{figure}
\includegraphics[width=0.35\linewidth]{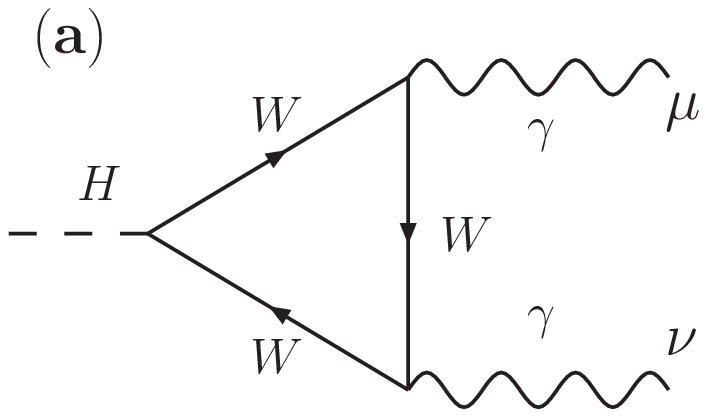}\;
\includegraphics[width=0.35\linewidth]{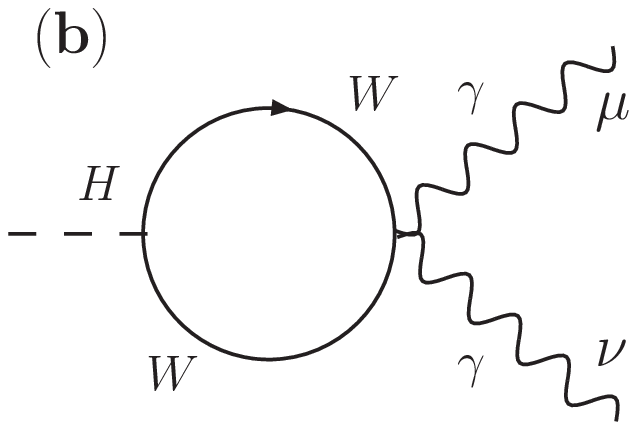}\\
\includegraphics[width=0.22\linewidth]{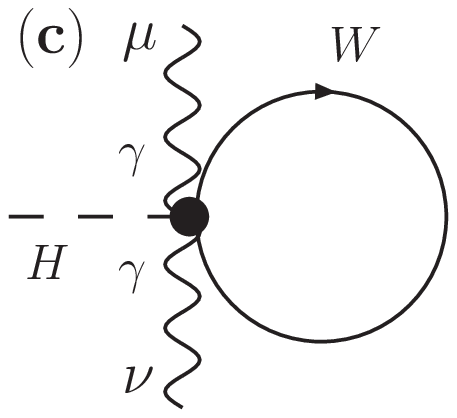} \;
\includegraphics[width=0.35\linewidth]{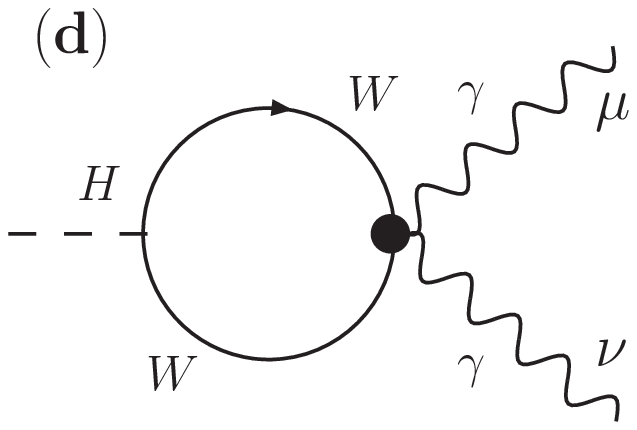}\\
\includegraphics[width=0.32\linewidth]{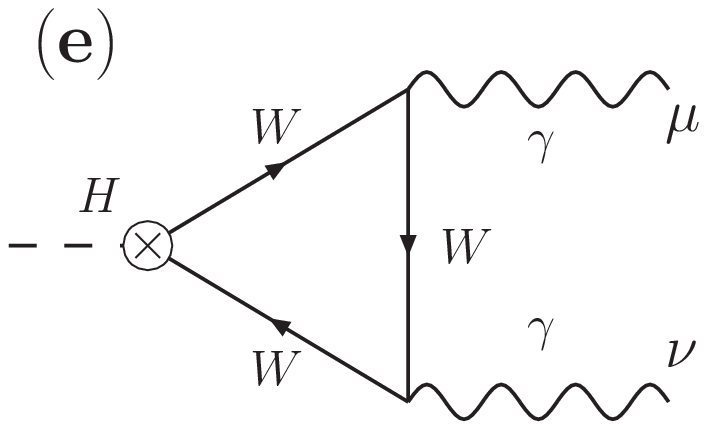}
\includegraphics[width=0.32\linewidth]{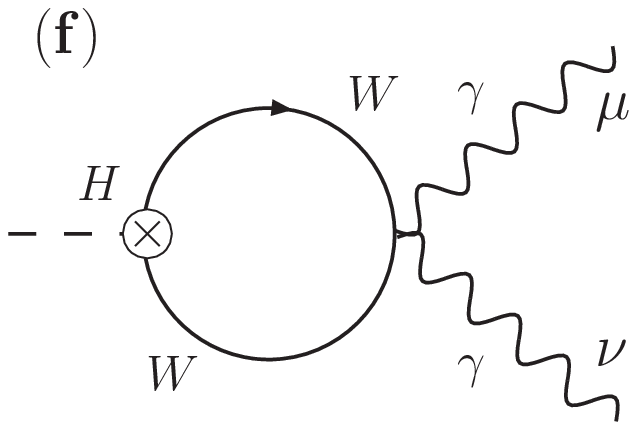}
\includegraphics[width=0.32\linewidth]{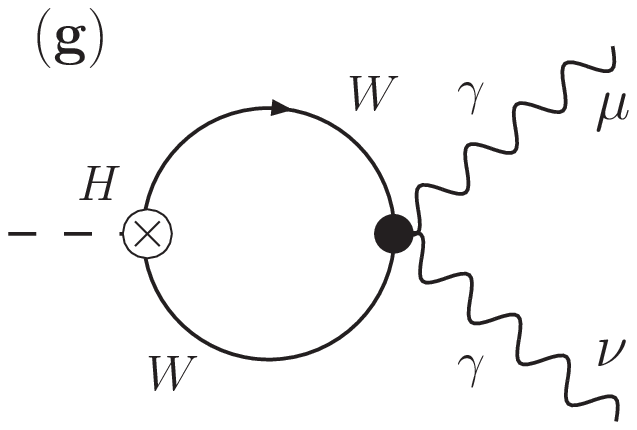}
\caption{\label{fig:Fdiags}Feynman diagrams (see text for description).}
\end{figure}

{\it Summary of diagrams} --- In our unitary-gauge lattice-regulated
calculation there are seven Feynman diagrams for the decay process, shown in
Fig.~\ref{fig:Fdiags}. In the triangle (a) and turnip (b)
diagrams, the vector boson vertices arise from the pure $SU(2)$ gauge vector
boson action and are similar to ones that arise in QCD. 
The ankh (c) and lattice turnip (d) diagrams arise from the additional lattice
Feynman rules. Diagrams (e), (f) and (g) are modifications of (a), (b) and
(d) to include the lattice forward derivative $\nabla_{\mu}^{+}$ coupling due
to the $(\nabla_0^{(+)}H)W^+W^-$ term in the action. Note that there is no
such modification of the ankh diagram as it would require the photons to have
Lorentz index $\mu=0$, which they do not.

{\it Cancellation of divergences} --- 
The result of combining these diagrams must be finite since there is no
tree-level process for $H \to \gamma\gamma$, so no counter-term can be
constructed to cancel any remaining divergences through the usual
renormalization prescription.  Therefore divergences $(aM_W)^{-n}$ present in
individual diagrams must cancel. To investigate this cancellation in more
detail, we fix $\tau$ and study the dependence on $aM_W$. By verifying the
cancellation of divergences and obtaining a finite answer for the matrix
element, ${\cal M}_{11}$, we perform a very strong check on the calculation.
Diagram (a) potentially has a $(aM_W)^{-6}$ divergence but this vanishes
trivially. Both (a) and (b) have $(aM_W)^{-2}$ and $(aM_W)^{-4}$
divergences. The $(aM_W)^{-4}$ divergences must cancel between the two
diagrams, since no such divergences appear in any other diagram in
Fig.~\ref{fig:Fdiags}. This is confirmed numerically, and we also find that
the $(aM_W)^{-2}$ divergence cancels between these two diagrams.
The lattice diagrams (c) and (d) contain $(aM_W)^{-2}$ divergences which
cancel between them. This is a good check for the derived lattice Feynman
rules.
The derivative diagrams (e)-(g) contain modifications of the $HW^+W^-$ vertex
which restricts the Lorentz index on the $W$ bosons at this vertex to be
$\mu=0$. They each clearly contain a $(aM_W)^{-2}$ divergence, and we find
that the contributions from all three diagrams are needed for
cancellation. This acts as another strong check on the lattice Feynman rules
since the contribution from (g) is needed to cancel the divergences
of (e) and (f), thus linking the rule generating diagram (d) with the standard
lattice rules which give rise to diagrams (a) and (b). Indeed, we find
numerically that diagrams (e), (f) and (g) cancel exactly.

\begin{figure}[h!]
\begin{center}
\includegraphics[width=0.85\columnwidth,clip]{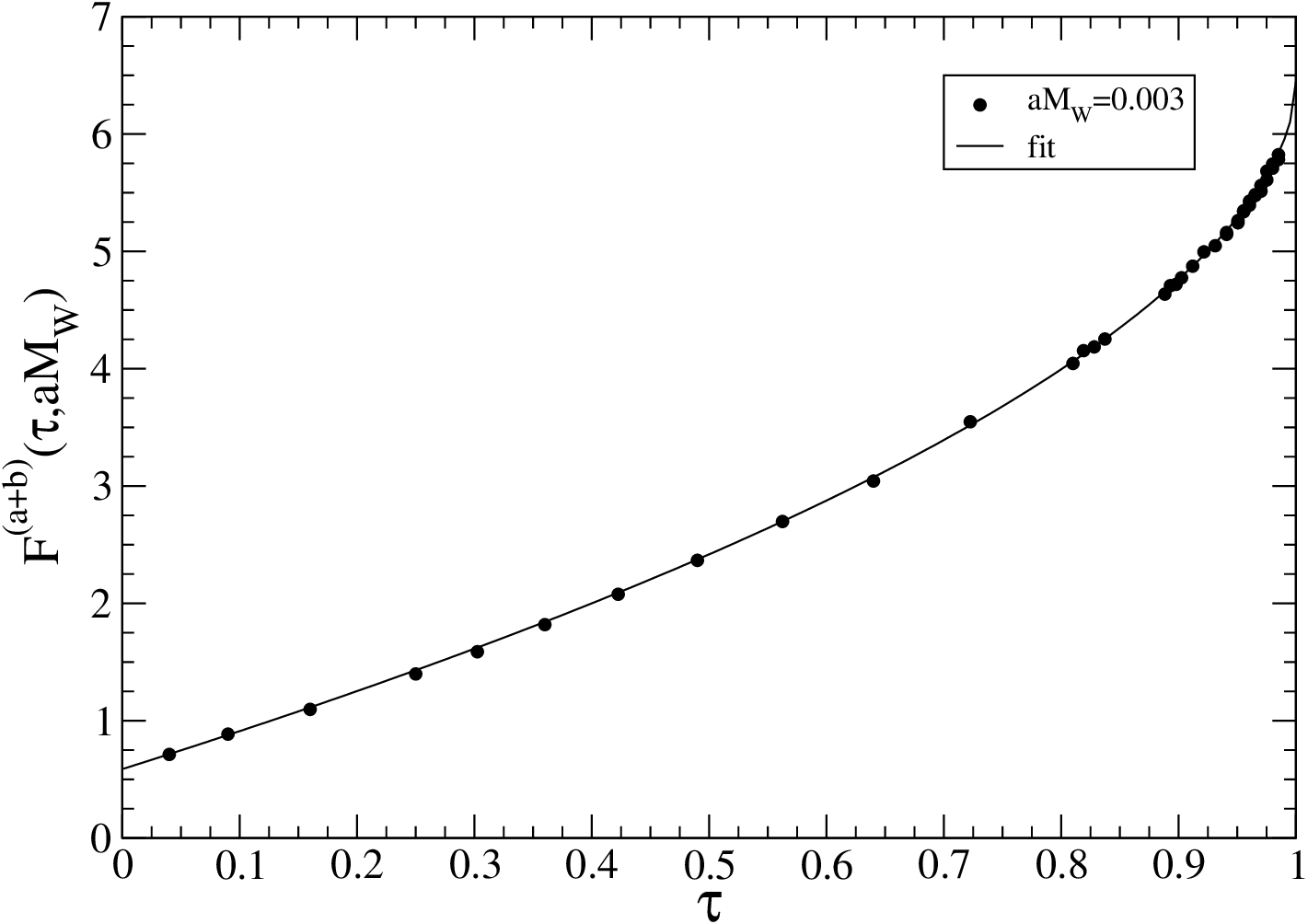}
\end{center}
\vspace{-5mm}
\caption{\label{higgsab} A plot for illustration of $F^{(a+b)}(\tau,aM_W)$ for
  $aM_W = 0.003$ and the fit from Eq.~(\ref{fitab}) for $F^{(a+b)}(\tau)$.
  There are no observable lattice artifact effects for any value of $aM_W$
  used.}
\vskip 5 truemm
\begin{center}
\includegraphics[width=0.85\columnwidth,clip]{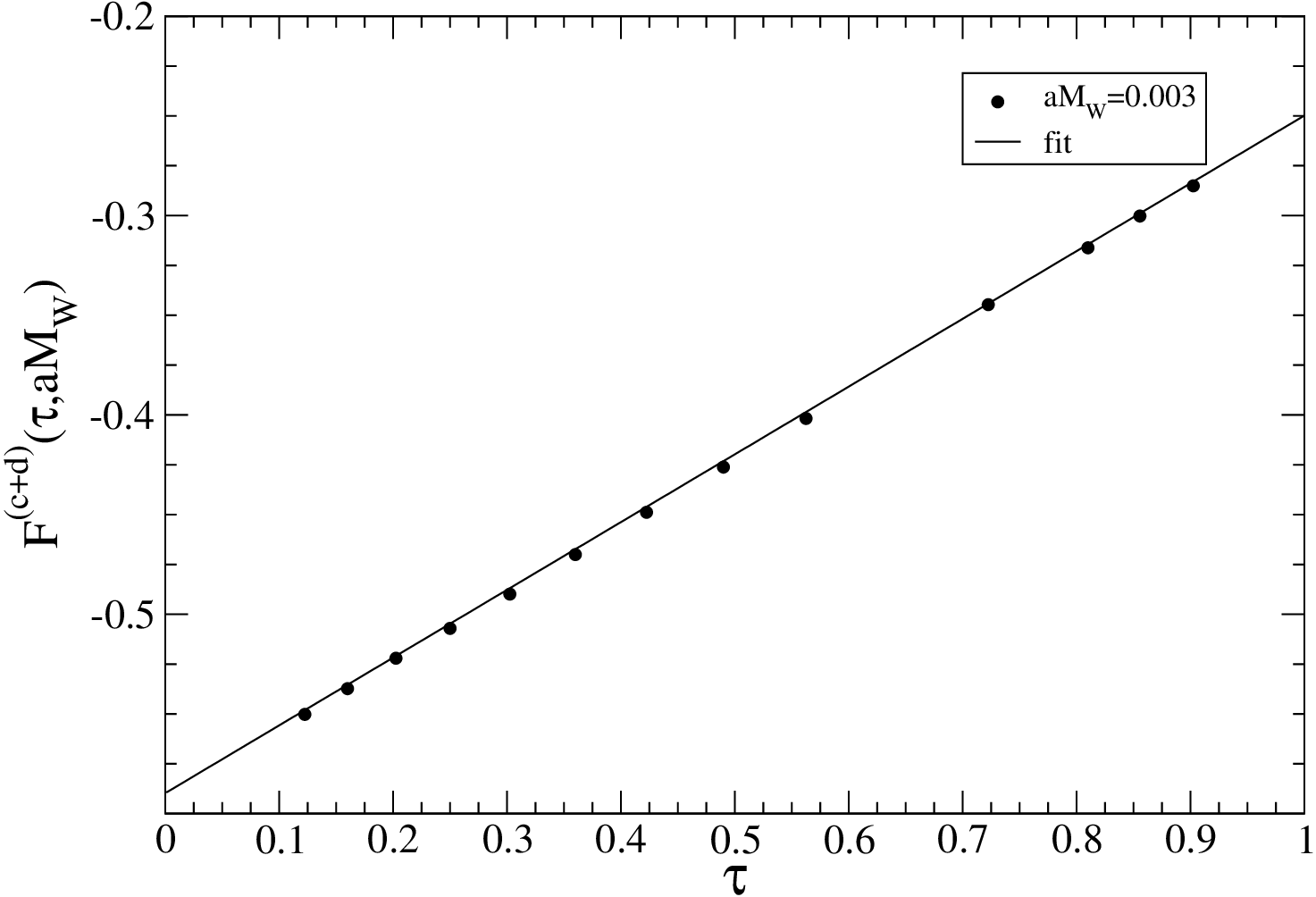}
\end{center}
\vspace{-5mm}
\caption{\label{higgscd} A plot for illustration of $F^{(c+d)}(\tau,aM_W)$ for
  $aM_W = 0.003$ and the fit from Eq.~(\ref{fitcd}) for $F^{(c+d)}(\tau)$.
  While there is a lattice artifact term in the fit, its effect even for the
  largest vale of $aM_W$ used is not discernible on the plot being a
  correction of order 1\%}
\end{figure}

{\it $\tau$ dependence} --- We deduce $F(\tau)$ by performing a simultaneous
fit in $(\tau, aM_W)$ to $F(\tau,aM_W)$ so that the continuum extrapolation $a
\to 0$ can be done. We initially fit the sum of the contributions from
diagrams (a)+(b) and from (c)+(d) separately, denoting these $F^{(a+b)}$ and
$F^{(c+d)}$, respectively. For each fixed $\tau$ value we compute
$F(\tau,aM_W)$ in both cases for values of $aM_W$ $0.003 \le aM_W \le
0.2$. We find that $F^{(a+b)}(\tau,aM_W)$ has no discernible $aM_W$
dependence in this range at fixed $\tau$ and we find the fit result
\begin{equation}
F^{(a+b)} = 2.088(8) + 0.660(17)\tau+1.500(3)\left(2-\frac{1}{\tau}\right)[\arcsin{\sqrt{\tau}}]^2 \label{fitab}
\end{equation}
with $\chi^2=0.25$. There is no improvement in fit from including lattice
artifact $aM_W$ polynomials or logarithms.  For illustration, we show in
Fig.~\ref{higgsab} the data for the smallest $aM_W$ value, $aM_W = 0.003$, and
the best fit curve in (\ref{fitab}).

The contribution of the lattice-induced diagrams, (c) and (d), does have a
dependence on lattice artifacts and is well-described by the function
\begin{equation}
F^{(c+d)} = -0.590(1)+0.340(1)\tau-0.24(3)(aM_W)^2\log{aM_W}\;, \label{fitcd}
\end{equation}
with $\chi^2=0.65$.  The fits indicate that there is no need to include a
polynomial in $aM_W$ and that the lattice artifact term turns out to be
independent of $\tau$. In Fig.~\ref{higgscd} we show the data for the smallest
$aM_W$ value, $aM_W = 0.003$, and the best fit curve in Eq.~(\ref{fitcd}). For
the values of $aM_W$ used we do not illustrate the lattice artifact dependence
by plotting the data for other values of $aM_W$ since for the largest value
used the correction is of order 1\% and unresolvable on the scale of the graph
shown.  The well-controlled continuum extrapolation is, to a high degree, due
to the use of the Symanzik improved gauge action. The derivative diagrams,
(e)-(g), cancel exactly, and give no net contribution.  For the graphs shown
the error bars are smaller than the symbol size.

From Eq.~(\ref{fitab}) and (\ref{fitcd}) we see that $F^{(a+b)}(\tau)$ 
contains the arcsin term which gives rise to the threshold behaviour, as we 
would expect, and the additional lattice terms in $F^{(c+d)}(\tau)$ give a 
simple linear behaviour in $\tau$ which is, however, vital to the question 
in hand namely the value of $c_2$. We note also that neither 
$F^{(a+b)}(\tau)$ or $F^{(c+d)}(\tau)$ individually vanish at $\tau = 0$.

A fit to the contributions of all four diagrams (a) - (d) gives the final result in the
continuum limit of
\begin{equation}
F(\tau) = 1.498(8)+1.000(17)\tau+1.500(3)\left(2-\frac{1}{\tau}\right)[\arcsin{\sqrt{\tau}}]^2\;,
\end{equation}
which is in strong agreement with the established dimensional regularization
result.  We note the fact that $F(0) = 0$ arises directly from our
calculation, as it does in dimensional regularization, without recourse to the
Dyson subtraction of other calculations in $D=4$
\cite{Gastmans:2011ks,*Gastmans:2011wh}.

{\it Conclusions} --- 
Physical predictions should not depend on the particular way a 
quantum field theory is regulated.  This paper shows how perturbative lattice calculations can be used to check the correctness of continuum calculations of observables, so long as the questions involved do not directly involve couplings of gauge fields to complex-representation Weyl fermions.  Agreement between the lattice
calculation here of the $W$-mediated contribution to $H\to\gamma\gamma$
and the longstanding result obtained with dimensional regularization
should put to rest any controversy regarding the Standard Model
prediction for this decay channel.  

{\it Acknowledgments} --- We are very grateful to B.~C.~Allanach, T.~Ishii, 
and S.~Lee for helpful discussions, and especially to
A.~Maharana for collaboration in the early stages. 
This work was supported by STFC under grant ST/G000581/1. The calculations
for this work were, in part, performed on the University of Cambridge HPCs as
a component of the DiRAC facility jointly funded by STFC and the Large
Facilities Capital Fund of BIS. 

\bibliographystyle{apsrev4-1}
\bibliography{HiggsTo2Gamma} 

\end{document}